\documentclass[conference]{IEEEtran}
\usepackage{amsmath,amssymb,amsfonts}
\usepackage{algorithmic}
\usepackage{graphicx}
\usepackage{textcomp}
\usepackage{xcolor}
\usepackage{hyperref}

\begin{document}

\title{Towards Stateless Clients in Ethereum: Benchmarking Verkle Trees and Binary Merkle Trees with SNARKs}

\author{\IEEEauthorblockN{Jan Oberst}
\IEEEauthorblockA{\textit{Institute of Information Security and Dependability (KASTEL)}\\
\textit{Karlsruhe Institute of Technology (KIT)}\\
Karlsruhe, Germany \\
jan.oberst@student.kit.edu}}

\maketitle

\begin{abstract}
Ethereum, the leading platform for decentralized applications, faces challenges in maintaining decentralization due to the significant hardware requirements for validators to store Ethereum's entire state. To address this, the concept of stateless clients is under exploration, enabling validators to verify transactions using cryptographic witnesses rather than the full state. This paper compares two approaches currently being discussed for achieving statelessness: Verkle trees utilizing vector commitments and binary Merkle trees combined with SNARKs. Benchmarks are performed to evaluate proving time, witness size, and verification time. The results reveal that the Verkle tree implementation used for benchmarking offers proving and verification times on the order of seconds and proof sizes on the order of one MB. The SNARK-based Merkle trees exhibit slow proof generation times, while offering constant and fast verification time. Overall, the results indicate for Verkle trees to provide a more practical solution for Ethereum’s stateless future, but both methods offer valuable insights into reducing the state burden on Ethereum nodes. We make the code used for benchmarking available on \href{https://github.com/JanIsHacking/verkle-vs-binary-bench}{GitHub}.
\end{abstract}


\section{Introduction}
Since its launch in 2015, Ethereum has become the leading platform for decentralized applications \cite{wu2021first}. Built on a blockchain that supports smart contracts, Ethereum has revolutionized how applications are built and executed in a decentralized manner. Having transitioned from a Proof of Work to a Proof of Stake consensus mechanism, Ethereum has separated block producers from consensus nodes. The latter vote on the validity of a block for the network to reach consensus on the current state of the blockchain. With consensus being a core mechanism for the security of Ethereum, decentralization of consensus nodes is very much desirable. However, as all nodes currently are required to store the entire state, decentralization is hindered by high hardware requirements such as access to significantly large, fast memory.

To address this issue, the implementation of \emph{weak statelessness} is included in the current Ethereum roadmap \cite{stateless_eth_roadmap}. Here, weak statelessness refers to a consensus node not being required to store the entire state to attest on proposed blocks. Instead, a witness of the validity of the block is included in the block itself which can be used by the consensus node---and anyone else interested---to verify the block's correctness.

Currently, Ethereum's state is stored in a Merkle Patricia Trie (MPT) \cite{wood2014ethereum}. The hashes of all account states in Ethereum make up the leaves of the trie and inner nodes hold the hash of their children nodes. Two MPTs therefore represent the same state if they have the same root hash. Furthermore, a proof about a specific leaf being included in the trie at a given position (e.g. the specific state of an account) can be constructed by providing all the sibling nodes on the path from the leaf to the root node. Such a Merkle proof can be verified by hashing the leaf data and iteratively calculating the values of the inner nodes using the provided sibling nodes all the way from the leaf to the root. Comparing the resulting root hash and the root hash referenced in the block concludes the proof. The list of sibling nodes therefore represents a witness for the proof that said leaf is included in the state tree at the said position. 

To provide a witness of the correct transaction execution of all transactions in one block, a naive approach would be to provide such a witness for every account involved in a transaction, before and after executing the transactions. This ensures that the states of all accounts involved in transactions have changed correctly. All accounts not involved in a transaction are part of a subtree whose root is a sibling node in some Merkle proof. Changes of uninvolved account states can therefore be detected by checking whether sibling nodes that represent such a root don't change between the states of before and after applying the transactions. In conclusion, the collection of witnesses for the Merkle proofs forms one overall witness attesting to the change in state being in accordance to the transactions included in the block.

This naive approach has a size complexity of

\begin{equation}
    O(T B k \log_{k} N)
\end{equation}
\label{eq:witness-size-hexary}

where $T$ is the number of transactions per block (TPB), $B$ is the size of an inner node in the tree, $k$ is the arity of the tree, and $N$ is the number of accounts. Note that the size scales linearly with $k$, as all sibling nodes have to be provided for all the levels $\log_{k} N$. With a few thousand TPB, 256-bit hashes, the MPTs being hexary trees, and hundreds of millions of accounts, a block witness is in the tens of Megabytes (MBs) large. Including a witness of this size into a block is therefore infeasible.

This motivates exploring alternative data structures and proof compression techniques to reduce the witness size. The main avenue currently being followed by the Ethereum core developers are Verkle trees, first introduced by Kuszmaul in \cite{kuszmaul2019verkle}. The idea of applying Verkle trees in Ethereum was developed by Buterin in \cite{vitalik2021verkle} and consolidated in Ethereum Improvement Proposal 6800 \cite{buterin2023eip6800}. The use of vector commitments removes the need to provide sibling nodes in the witness and a constant size proof can be generated over the internal nodes due to the additively homomorphic properties of polynomial commitments. The structure of Verkle trees, what a witness looks like, and theoretical estimates of the witness sizes are described in Section \ref{sec:verkle-trees}.

Another approach investigated in this work is the use of binary Merkle trees. Switching to a unified binary Merkle tree for Ethereum's state representation was first introduced by Ballet and Buterin in \cite{eip3102}. In \cite{vitalik2021verkle}, Buterin touches on using Succinct Non-Interactive Arguments of Knowledge (SNARKs) to reduce witness sizes and allow for stateless clients. With the size of the used SNARKs being constant, the size complexity of the block witness is reduced to $O(T)$. This results in a witness only a few MBs large. Details on the generation of the witnesses and the size estimation are described in Section \ref{sec:applying-snarks-to-binary}.

The goal of this work is twofold: 

\begin{enumerate}
    \item Consolidate the knowledge on Verkle trees and binary Merkle trees with SNARKs for enabling stateless clients in Ethereum.
    \item Provide benchmark results that contribute to informed decisions for future developments.
\end{enumerate}

In the first step, related work on benchmarking Verkle trees and binary Merkle trees with SNARKs is presented in Section \ref{sec:related-work}, along with an explanation of their fundamentals in Section \ref{sec:fundamentals}. Subsequently, the methodology and implementation of the benchmarks are outlined in Section \ref{sec:methodology} and \ref{sec:implementation} respectively. The results of the benchmarks measuring proving times, witness sizes, and verification times for both approaches are presented in Section \ref{sec:results}. In Section \ref{sec:discussion}, the results and their implications are discussed along with their limitations. Finally, a conclusion is drawn and an outlook on future work is presented in Section \ref{sec:conclusion}.

\section{Related Work}
\label{sec:related-work}
The literature on the application of both Verkle trees and binary trees with SNARKs for state representation in Ethereum is very sparse. Relevant publications are divided by their relationship to the respective approach and presented in the following.

\textbf{Verkle Trees.} Our literature review could not reveal any publications benchmarking Verkle trees. While there exist a handful of implementations of Verkle trees (e.g. \cite{go_verkle}, \cite{verkle-tree-intmax}, \cite{verkle-trie-drasutis}) written in Go and Rust, none of them have published benchmarking results. The only public results could be found in a benchmarking framework by Hagopian \cite{vkt-proof-bench}, providing measurements on proving and verifying a variable number of keys given a tree of 1,000,000 leaves.

\textbf{Binary Trees with SNARKs.} To the best of our knowledge, the only publication that uses proof systems to reduce witness sizes for Merkle trees is \cite{kuznetsov2024enhanced}. The authors investigate the use of Scalable Transparent Arguments of Knowledge (STARKs) as well as SNARKs, and apply them to real and synthetic state data. For STARKs, measurements of proving times, witness sizes, and verification times are presented. For proofs using SNARKs, no measurements are provided, leaving a research gap.

Although El-Hajj and Roelink's work does not directly apply to Merkle proofs, it provides insights into proof systems \cite{el2024evaluating}. The authors benchmark different implementations of SNARK, STARK, and Bulletproof schemes across a varying number of MiMC hash evaluations. They find that while STARKs produce the largest proofs, they are the fastest to generate and verify.

\textbf{Research Questions.} Given the lack of benchmarks of Verkle trees and binary Merkle trees using SNARKs in the literature, the following research questions arise:

\begin{enumerate}
    \item How do existing implementations of Verkle trees and binary trees with SNARKs perform in terms of proving time, witness size, and verification time?
    \item How do both approaches compare?
\end{enumerate}

The research questions are addressed in this work by providing a benchmarking framework for both approaches and running the benchmarks for different tree sizes on two different machines.

\section{Fundamentals}
\label{sec:fundamentals}
The following section assumes general familiarity with the Ethereum protocol. For an introduction to Merkle trees in the context of blockchain systems, refer to \cite{liu2021merkle}.

First, the structure of Verkle trees is introduced along with a witness size estimation. An overview of SNARK proof schemes is given subsequently and the Permutations over Lagrange-bases for Oecumenical Noninteractive arguments of Knowledge (PLONK) scheme is described in more detail. A witness size estimation applying SNARKs to binary Merkle trees is provided as well. For the witness size estimates, a worst case of 5,000 keys to prove per block is assumed.

\subsection{Verkle Trees}
\label{sec:verkle-trees}
Verkle trees are a cryptographic data structure designed to optimize the storage and witness sizes of Ethereum's state, serving as a more efficient alternative to MPTs. They provide a trade-off between witness size and computational overhead for witness generation. Like Merkle trees, Verkle trees structure data hierarchically, allowing for compact proofs about the inclusion of specific data in the tree. However, instead of relying solely on cryptographic hash functions, Verkle trees leverage polynomial commitment schemes to achieve much smaller witness sizes, making them highly attractive for stateless clients.

At a high level, Verkle trees are similar to Merkle trees in that they allow the construction of a proof of inclusion for a given piece of data by traversing the tree from the leaf node (where the data resides) to the root node. However, unlike Merkle trees, which use hash functions to commit to data at each level, Verkle trees use vector commitments. Vector commitments are a type of cryptographic primitive that allow the commitment to a large vector of data (such as a tree's child nodes) while enabling efficient proofs about any individual element within that vector. Polynomial commitments, a type of vector commitment, enable the prover to create compact, constant-sized proofs that allow for the verification of multiple tree nodes' integrity.

The core advantage of Verkle trees lies in the ability to aggregate multiple proofs into a single, compact multiproof, which significantly reduces the size of the witness required for validation \cite{feist-multiproof}. In contrast to the logarithmic-size Merkle proofs, the witness sizes in Verkle trees are therefore much smaller and independent of the tree depth.

\textbf{Verkle Witness Size.} A Verkle witness consists of three main components \cite{hagopian-verkle-proof}:
\begin{enumerate}
    \item The leaves to be proven, representing the specific data entries in the state.
    \item The intermediate vector commitments from the leaf to the root, representing the paths in the tree.
    \item A multiproof over all parent-child relationships between the intermediate node commitments, ensuring the integrity of the proof.
\end{enumerate}

The witness size for Verkle trees can be estimated by considering the size of each component. A key benefit of Verkle trees is that only the commitments, rather than all the siblings on the paths, need to be provided, drastically reducing the overall witness size.

The total witness size for an Ethereum block can be estimated using the following equation:
\begin{equation}
    \resizebox{\columnwidth}{!}{$
        2 \times 5000 \times 32 \text{ Bytes} + 10000 \times 48 \text{ Bytes} + 200 \text{ Bytes} = \sim 0.8 \text{ MBs}
    $}
\end{equation}

Each part of this equation corresponds to a different component of the witness:
\begin{itemize}
    \item $2 \times 5000 \times 32$ Bytes: This term accounts for the leaves to be proven. Each leaf consists of a key and a value of size 32 Bytes.
    \item $10000 \times 48$ Bytes: This term represents the intermediate node commitments that must be included in the proof, assuming a tree with $2^{32}$ leaves. On the first level is the root. On the second level, all 256 commitments will be included. On the third level is a bit of redundancy in the commitments and on the fourth level are the 5,000 leaves, equating to an estimate of 10,000 intermediate commitments. Each vector commitment is 48 Bytes in size.
    \item $200$ Bytes: The multiproof ensuring the correctness of all parent-child relationships between the intermediate commitments is of constant size.
\end{itemize}

In total, this gives a witness size of approximately 0.8 MBs per block. This is a significant improvement over the tens of MBs required by MPTs.

\subsection{SNARKs}
Succinct Non-interactive Arguments of Knowledge (SNARKs) are a powerful cryptographic tool that enables the creation of short, efficient proofs for complex computations. A SNARK allows a prover to convince a verifier that they know the result of an arbitrary computation, without revealing the details of the computation itself. The main feature of SNARKs is that the resulting proof size is succinct, meaning the proof is at least exponentially smaller than the computation being proven, while still being efficiently verifiable.

At the core of SNARKs is the concept of arithmetic circuits. Instead of directly encoding a program or computation, SNARKs require the computation to be expressed as a series of simple arithmetic operations such as addition and multiplication over a finite field. These operations are arranged into an arithmetic circuit, where each gate in the circuit performs one of the above mentioned operations. The prover has to transform the calculation they want to prove into an arithmetic circuit. Given this circuit, the prover can generate a SNARK that convinces the verifier that they know the correct output of a computation, which is represented by the arithmetic circuit. The prover provides the input and output of the computation to the verifier, along with public parameters that the SNARK scheme participants generated in a so-called setup ceremony.

The witness in a SNARK therefore consists of two components: (1) the inputs, which represent the public data required for verification (such as the leaves and Merkle roots in state proofs), and (2) the proof, which ensures the computation was performed correctly. The witness enables the verifier to quickly check whether the result of the computation is correct, without needing to re-execute the computation itself.

\textbf{The PLONK SNARK Scheme. }PLONK is a state-of-the-art SNARK scheme \cite{gabizon2019plonk}. One of its key features is its updatable, universal trusted setup, which makes it reusable across different computations. Unlike earlier SNARKs that required a trusted setup for every arithmetic circuit, PLONK only requires one universal trusted setup. This setup can then be used for multiple computations without needing to be repeated.

Another important property of PLONK is that it supports efficient proving and constant-sized proofs, regardless of the size or complexity of the underlying computation. This is made possible through its use of custom gates and permutation arguments, which allow the SNARK to be applied to more general computations. Furthermore, PLONK achieves constant verification time, meaning that no matter how large the circuit is, the time required for verification remains fixed.

\subsection{Applying SNARKs to Binary Merkle Tree Proofs}
\label{sec:applying-snarks-to-binary}

Binary Merkle trees are widely used to represent and verify the state of data structures in blockchain systems \cite{liu2021merkle}. When using SNARKs for proving Merkle branches, the structure of the Merkle branch as well as the computation of the hash values themselves need to be encoded into arithmetic circuits.

To prove the inclusion of $N$ keys in a binary Merkle tree, one usually must provide $N$ Merkle proofs, leading to unfeasibly large witness sizes. SNARKs offer an elegant solution to this problem by allowing the prover to generate a single, succinct proof for each Merkle branch. In this approach, each Merkle branch is transformed into its own arithmetic circuit as described earlier, which the prover can then use to generate the SNARK. The verifier no longer needs to verify each individual branch separately but can instead verify the individual SNARK proofs, which attest to the correctness of each branch.

The witness size for proving 5,000 Merkle branches with SNARKs can be estimated as: \begin{equation} 
    2 \times 5000 \times (192 + 2 \times 32) \text{ Bytes} = \sim 2.6 \text{ MBs}
\end{equation}

The estimated witness consists of $2 \times 5000$ Merkle branches per block, one for each leaf to prove before and after the transaction executions. Each Merkle proof witness consists of the SNARK, 192 Bytes in size, and the leaf's key-value pair.

This yields a total witness size of approximately 2.6 MBs, including the leaf data and the SNARK proofs. Although larger than the witness for Verkle trees, this SNARK-based proof allows each of the 5,000 Merkle branches to be verified with individual, constant-sized SNARK proofs, regardless of the tree depth.

\section{Methodology}
\label{sec:methodology}
To assess the potential of Verkle trees and binary Merkle trees for Ethereum state representation and their suitability for enabling stateless clients, the benchmarks presented in this work focus on three key metrics: proving time, proof size, and verification time. Both Verkle and binary Merkle trees were evaluated across varying tree sizes, starting at $2^5$ leaves and increasing them until either $2^{32}$ is reached or benchmarking becomes infeasible due to the system running out of main memory. Starting at $2^{14}$, every second power of two is used for the number of leaves in the tree. For each tree size, the metrics are averaged over ten runs to account for the possible variance in runtime. Additionally, the number of keys to prove per tree is set to 5,000. For trees with less than 5,000 keys, all of them are proven.

The benchmarking experiments were conducted on two different machines to account for hardware variability. Their specifications are displayed in Table \ref{tab:machines-specs}.

\begin{table}[tbp]
\caption{The two machines on which the benchmarks were performed.}
\begin{center}
\begin{tabular}{|c|c|c|c|}
\hline
\textbf{\#} & \textbf{Operating System} & \textbf{CPU} & \textbf{RAM} \\
\hline
1 & Windows 10 22H2 & Intel i5-4690K & 22 GiB \\
\hline
2 & Ubuntu 23.10 & AMD Ryzen 5975WX 32-Core & 125 GiB \\
\hline
\end{tabular}
\label{tab:machines-specs}
\end{center}
\end{table}

\section{Implementation}
\label{sec:implementation}
The implementation was done in Rust, making use of existing libraries for the required data structures and proving schemes.

For the Verkle trees, the implementation of the tree was taken from the Rust library "rust-verkle" \cite{rust_verkle}. It is worth noting that this implementation relies on main memory and does not store any data on disc. The interface of the tree was augmented to allow for proof size measurements and was integrated into the benchmarking framework measuring the metrics described above.

For the binary Merkle trees, the tree implementation using the Poseidon hash function was taken from the "poseidon-merkle" library \cite{poseidon_merkle}. The implementation was adapted to accommodate a tree arity of two. 

For the SNARK-based proof system, we utilized the PLONK SNARK scheme from the "dusk-plonk" library \cite{plonk_dusk}. This provided an efficient mechanism for generating and verifying SNARK-based proofs. As with the Verkle trees, the aforementioned data structures and schemes were integrated into the benchmarking framework presented in this work to align with the methodology described in Section \ref{sec:methodology}. 

None of the libraries mentioned above, including the one for Verkle trees, are parallelized. The benchmarking runs for Verkle trees are parallelized, while parallelizing the binary Merkle tree benchmarks was not possible without modifying the library's code. The code used for the benchmarks, along with all measurements and a short guide on how to run them, is publicly available on GitHub \cite{verkle_vs_binary_bench}.

\section{Results}
\label{sec:results}
The measured proving times, witness sizes, and verification times for Verkle trees and binary trees with SNARKs on the machines listed in Table \ref{tab:machines-specs} are displayed in Figures \ref{fig:results-verkle} and \ref{fig:results-binary}.

\subsection{Verkle Trees}

\begin{figure}[tbp]
\centering
\includegraphics[width=\columnwidth]{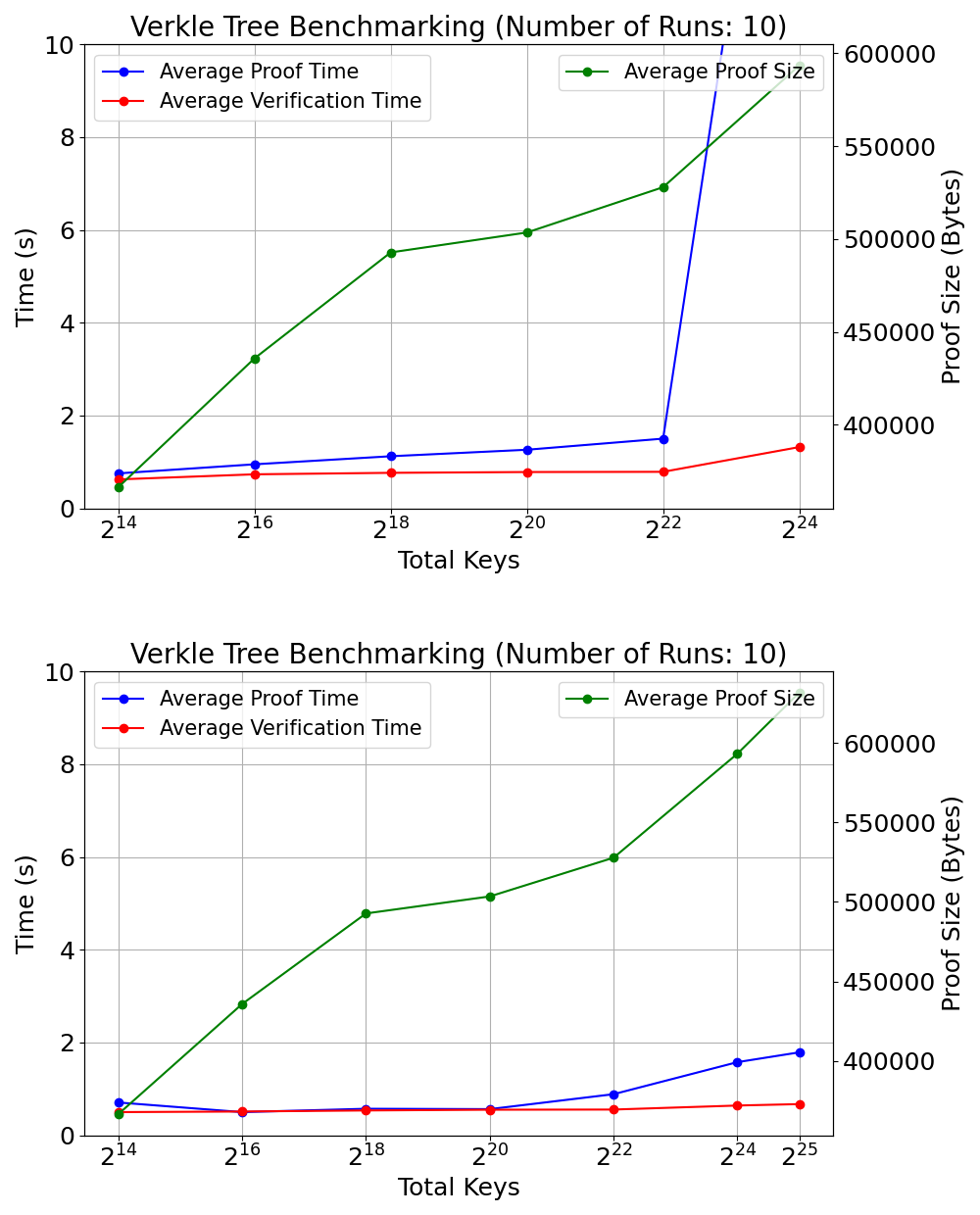}
\caption{Measured proving times, proof sizes, and verification times for Verkle trees on Machine 1 (top) and Machine 2 (bottom) averaged over ten runs.}
\label{fig:results-verkle}
\end{figure}

For better readability, only the measurements for trees of size $2^{14}$ and larger are displayed in the graphs. The maximum number of leaves measured are $2^{24}$ and $2^{25}$ for Machines 1 and 2 respectively. The execution of benchmarks with larger tree sizes was killed by the operating system due to insufficient available main memory.

Generally, the measurements for Verkle trees are not surprising. As displayed in Figure \ref{fig:results-verkle}, the proof time grows slowly over the exponentially increasing number of keys in the trees. The witness sizes and verification times increase logarithmically with the number of keys in the tree. While the absolute values of proving and verification times are lower for Machine 2, the trends observable in the data are similar for both machines. However, it is worth noting that the high main memory usage for the largest trees leads to outliers in the proof generation time. This can possibly explained by the operating system writing data to the disc to free up main memory leading longer latencies.

\subsection{Binary Trees}
\begin{figure}[tbp]
\centering
\includegraphics[width=\columnwidth]{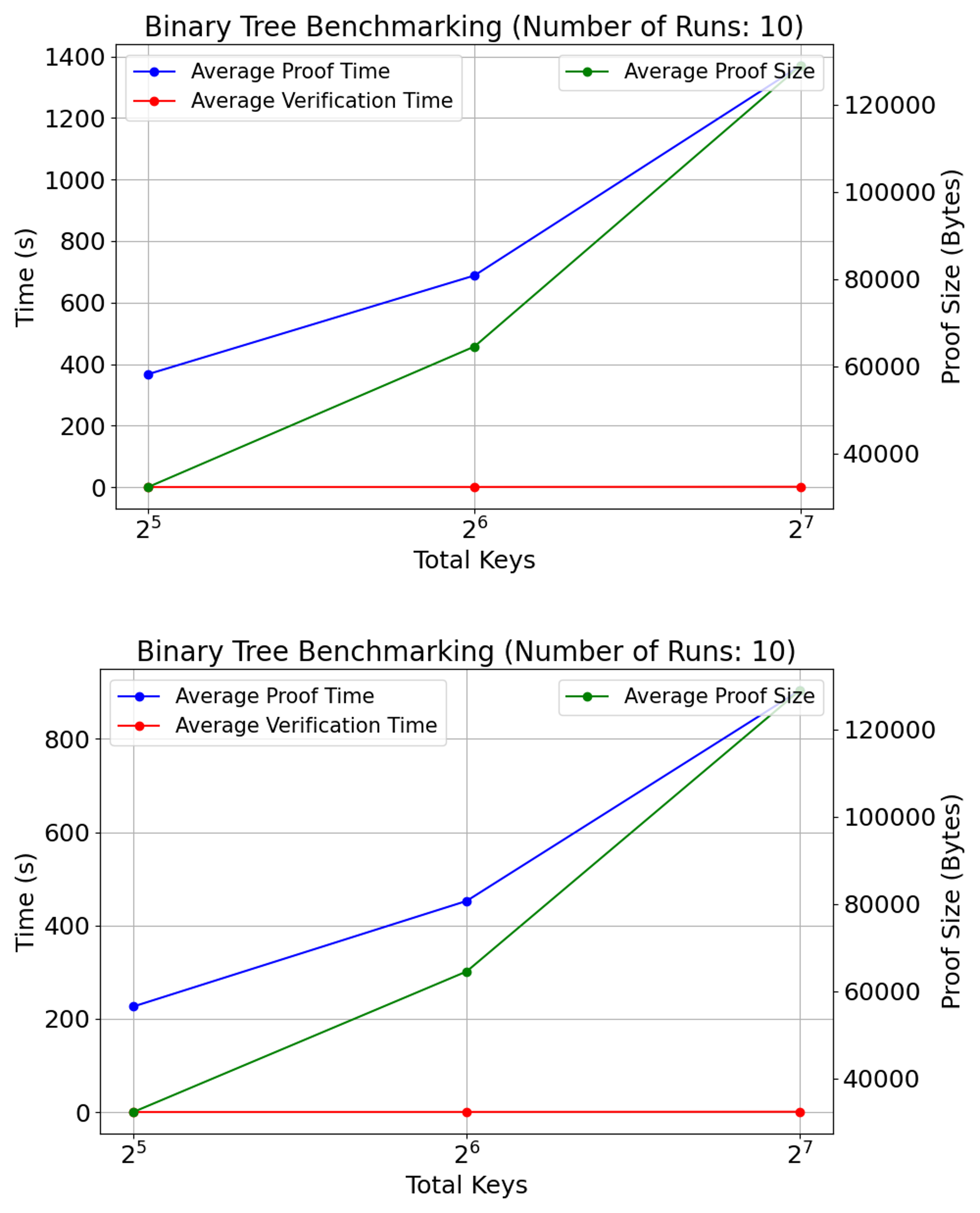}
\caption{Measured proving times, proof sizes, and verification times for binary trees with SNARKs on Machine 1 (top) and Machine 2 (bottom) averaged over ten runs.}
\label{fig:results-binary}
\end{figure}

The results of the binary tree benchmarks are displayed in Figure \ref{fig:results-binary}.

Proving takes several minutes on both machines for small tree sizes. The experiments were aborted for trees larger than $2^7$, as proving times of that length are not usable in practice. Compared to the proving times, the verification times are fast on the order of seconds. In general, SNARK proofs that utilize the PLONK scheme exhibit a constant size. The increase in proof size can be explained by the program always trying to prove 5,000 keys as explained in Section \ref{sec:methodology}. Hence, for trees with less than 5,000 leaves, all leaves are included in the proof. Consequently, the overall witness size is contingent upon the number of keys being proven, leading to an increase in size during benchmarks involving binary trees. For trees with 5,000 or more leaves, the proof size is expected to be constant.

\section{Discussion \& Limitations}
\label{sec:discussion}
The main limitation for drawing a conclusion on the comparison of Verkle trees and binary Merkle trees is the more optimized implementation of Verkle trees. While the results of the Verkle trees look promising for the application as state representation in Ethereum, the binary tree implementation does not scale due to its naive implementation. For each Merkle branch, a separate SNARK is generated, leading to higher witness sizes and more computational overhead.

As the tree sizes are also limited for Verkle trees due to their pure in-memory implementation, the results are not completely conclusive on the applicability of Verkle trees in Ethereum. The benchmarks could be extended with a database implementation of Verkle trees or by proving synthetic Verkle branches without explicitly storing the entire tree.

It is worth noting that for the absolute values of the proof generation time, the results on the more powerful Machine 2 are more conclusive, as the hardware requirements for execution clients would remain unchanged if Ethereum becomes stateless. In contrast, consensus clients should be able to verify the proofs on much less powerful hardware, hence the measurements on Machine 1 can be considered to be more meaningful.

Since the PLONK SNARK scheme is used for generating the proofs for the binary Merkle trees, a universal trusted setup between the participants is required. This can present a security risk to the proving scheme and must be evaluated carefully.

Both Verkle and binary tree state representations propose a unification of the account state tree and storage trees into one overall state tree. Therefore, determining the exact worst case number of key updates in the state tree is hard. However, we recognize that proving 5,000 keys per block may be an overestimate. The measurements presented in Section \ref{sec:results} are therefore likely to represent upper bounds to proving times, witness sizes, and verification times.

\section{Conclusion \& Outlook}
\label{sec:conclusion}
The results indicate that Verkle trees represent a viable choice for enabling stateless clients in Ethereum due to their scalability, small witness sizes, and fast verification times. However, no definitive conclusions can be drawn about the practicality of binary Merkle trees with SNARKs, as the implementation used in this study lacks necessary optimizations. The significantly longer proving times for binary trees may not accurately reflect their potential performance with further development. Therefore, while Verkle trees appear more suitable for realizing stateless clients in Ethereum, future work is required on the optimization of the implementation.

When considering an alternative tree structure to the current MPT, other performance metrics such as access and update times of tree data need to be taken into account. To fully grasp the effects of changing Ethereum's data structure for storing state, further research on this is required.

Both Verkle trees and SNARKs rely on elliptic curve cryptography, which is vulnerable to quantum attacks. As quantum computing is expected to advance significantly in the coming decade, long-term quantum-resistant alternatives need to be explored (e.g. STARKed Merkle proofs).

\section*{Acknowledgment}

We would like to thank Oliver Stengele, Jan Droll, and Prof. Hannes Hartenstein. Thank you for the introduction to the topic of Ethereum, your guidance along the way, and the valuable feedback after the presentations.

\bibliographystyle{plain} 
\bibliography{bibliography} 

\end{document}